\documentclass{ws-mpla}

\usepackage{graphicx} \usepackage{epsfig} \usepackage{amsmath,amssymb}

\begin{document}

\markboth{FRIB Theory Center Steering Committee} {FRIB Nuclear Theory}

\catchline{}{}{}{}{}

\title{Nuclear Theory and Science of the Facility for Rare Isotope Beams}


\author{\footnotesize A.B. BALANTEKIN} \address{Department of Physics,
  University of Wisconsin-Madison, Wisconsin 53706, USA}

\author{\footnotesize J. CARLSON} \address{Theoretical Division, Los
  Alamos National Laboratory, Los Alamos, New Mexico 87545, USA}

\author{\footnotesize D.J. DEAN} \address{Physics Division, Oak Ridge
  National Laboratory, Oak Ridge, Tennessee 37831, USA}

\author{\footnotesize G.M. FULLER} \address{Department of Physics,
  University of California, San Diego, La Jolla, California
  92093-0319}

\author{\footnotesize R.J. FURNSTAHL} \address{Department of Physics,
  The Ohio State University, Columbus, Ohio 43210, USA}

\author{\footnotesize M. HJORTH-JENSEN} \address{National
  Superconducting Cyclotron Laboratory and Department of Physics and
  Astronomy\\Michigan State University, East Lansing, Michigan 48824,
  USA}

\author{\footnotesize R.V.F. JANSSENS} \address{Physics Division,
  Argonne National Laboratory, Argonne, Illinois 60439, USA}

\author{\footnotesize BAO-AN LI} \address{Department of Physics \&
  Astronomy, Texas A\&M University-Commerce \\ Commerce, Texas 75429,
  USA}

\author{\footnotesize W. NAZAREWICZ} \address{Department of Physics \&
  Astronomy, University of Tennessee, Knoxville, Tennessee 37996,
  USA\\ Physics Division, Oak Ridge National Laboratory, Oak Ridge,
  Tennessee 37831, USA\\ witek@utk.edu}

\author{\footnotesize F.M. NUNES} \address{National Superconducting
  Cyclotron Laboratory and Department of Physics and Astronomy
  \\Michigan State University, East Lansing, Michigan 48824, USA}

\author{\footnotesize W.E. ORMAND} \address{Physics Division, Lawrence
  Livermore National Laboratory, Livermore, California 94551, USA}

\author{\footnotesize S. REDDY} \address{Institute for Nuclear Theory,
  University of Washington, Seattle, Washington 98195, USA}

\author{\footnotesize B.M. SHERRILL} \address{National Superconducting
  Cyclotron Laboratory and Department of Physics and Astronomy,
  \\Michigan State University, East Lansing, Michigan 48824, USA}

\maketitle

\pub{Received (Day Month Year)}{Revised (Day Month Year)}

\begin{abstract}
The Facility for Rare Isotope Beams (FRIB) will be a world-leading
laboratory for the study of nuclear structure, reactions and
astrophysics. Experiments with intense beams of rare isotopes produced
at FRIB will guide us toward a comprehensive description of nuclei,
elucidate the origin of the elements in the cosmos, help provide an
understanding of matter in neutron stars, and establish the scientific
foundation for innovative applications of nuclear science to society.
FRIB will be essential for gaining access to key regions of the
nuclear chart, where the measured nuclear properties will challenge
established concepts, and highlight shortcomings and needed
modifications to current theory.  Conversely, nuclear theory will play
a critical role in providing the intellectual framework for the
science at FRIB, and will provide invaluable guidance to FRIB's
experimental programs. This article overviews the broad scope of the
FRIB theory effort, which reaches beyond the traditional fields of
nuclear structure and reactions, and nuclear astrophysics, to explore
exciting interdisciplinary boundaries with other areas.

\keywords{Nuclear Structure and Reactions. Nuclear
  Astrophysics. Fundamental Interactions. High Performance
  Computing. Rare Isotopes. Radioactive Beams.}
\end{abstract}

\ccode{PACS Nos.: 21.60.-n, 21.30.-x, 21.65.-f, 24.10.-i, 24.80.+y,
  25.60.-t, 26.30.-k, 26.50.+x, 26.60.-c}

\section{Introduction: New Opportunities with FRIB}

Nuclear physics plays a key role in our quest to understand  the
Universe.\cite{Decadal2012,TribbleReport} In recent years, researchers
have made remarkable progress in our fundamental understanding of the
complex and fascinating system that is the
nucleus.\cite{Decadal2012,LRP07,RISAC} This progress has been driven
by new theoretical insights and increased computational power, as well
as by experimental access to new isotopes with a large excess of
neutrons or protons.  However, while much has been learned so far
about nuclear systems and associated phenomena, much remains to be
understood.

The Facility for Rare Isotope Beams (FRIB),\cite{FRIB} which is a
sponsored project of the US Department of Energy, Office of Science,
will push the frontiers of nuclear science by providing access to the
widest range of isotopes possible. For example, to explore changes in
shell structure, useful yields of nickel isotopes will be available
from $^{48}$Ni to $^{84}$Ni, thereby spanning the full range of
neutrons in the $pfg$-shells. The key is for the facility to deliver
very high power heavy-ion primary beams,\cite{Sym07} 400\,kW minimum, at
energies of at least 200\,MeV/u to be used in the production of rare isotopes.\cite{Mor04} These high-power beams will be
generated by a superconducting linear accelerator coupled with a
production area designed to operate at high current.\cite{Wei12,Wei13} The
broad scientific program requires rare-isotope beams at energies
ranging from stopped ions in traps\cite{Gee06} to ions at relativistic
energies of hundreds of MeV/u. FRIB will have these capabilities by
using full-energy beams following in-flight separation, stopped ions
thermalized using a variety of ion catcher schemes, and reaccelerated
ions delivered by the ReA superconducting linear
post-accelerator.\cite{Wei13} The goal is to reach reaccelerated beam energies of 12--20\,MeV/u.  As a result, the full complement of
direct reactions (including high-$\ell$ transfer and deep-inelastic
reactions) will be accessible for experimentation. An advantage of the
in-flight production and reacceleration approach is that isotopes of
all elements will normally be available with very short development
times and high efficiency (approaching 10 to 20\%). The technique will
also provide beams of most isotopes, even those with short (tens of
ms) half-lives. This offers the possibility to perform experiments,
for example, with beams of highly refractory elements along the
$N=126$ line of isotones below $^{208}$Pb, as required to improve
$r$-process nucleosynthesis models.

\begin{figure}[h!] 
  \centerline{\includegraphics[width=0.6\textwidth]{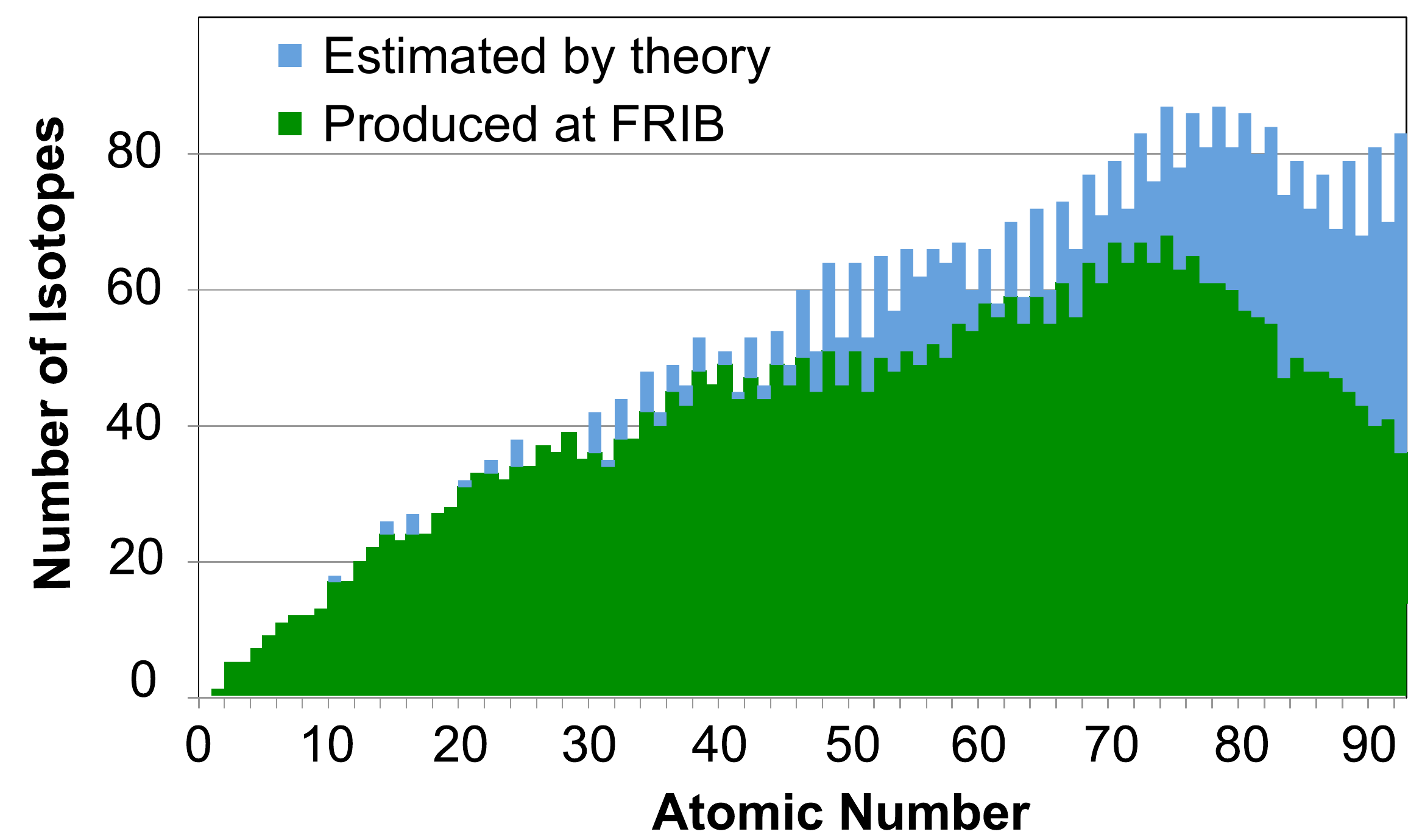}}
  \caption[T]{\label{FRIBrange} Number of isotopes of elements up to
    $Z = 92$ estimated to be produced in sufficient quantities at FRIB
    (green bars) to allow study of their structure and determine at
    least one property other than simple observation. The number of
    isotopes estimated to exist (blue bars) is taken from the recent
    theoretical survey of Ref.~\protect\refcite{(Erl12)}. FRIB is
    predicted to produce nearly 80\% of all possible isotopes in this
    range.}
\end{figure}

The anticipated range of isotopes to be available at FRIB, as
estimated using the LISE$^{++}$ program,\cite{FRIBrates} is shown in
Fig.~\ref{FRIBrange}.  It is compared to the predicted number of
possible isotopes from the average of the Density Functional Theory
(DFT) predictions.\cite{(Erl12)} With the 400\,kW beams of FRIB,
nearly 80\% of all isotopes of elements up to uranium may become
available for study. This includes many new isotopes estimated to lie
along the drip lines (perhaps even up to element $Z=61$ as shown in
the figure) and many nuclei with skins predicted to be greater than 0.5\,fm
(cf. Ref.~\refcite{(Kor13)}). In addition, the facility will have
provisions to collect unused isotopes and make them available for
experiments and applications in other fields, such as
medicine.\cite{Har12} One option being considered is to collect
isotopes produced in the beam dump by uranium stopping in water. This
approach would, for example, make nearly continuous supplies of
$^{223}$Rn or $^{225}$Ra available for fundamental interaction studies
searching for an atomic electric dipole moment.\cite{Eng13} Other
options are to provide a source of isotopes such as $^{67}$Cu or
$^{149}$Tb for medical studies.\cite{Har12}

In this broad overview of FRIB theory, we concentrate on major themes
pertaining to science, organization, and education aspects of the
effort. The references and links cited are not meant to be inclusive;
they should rather be used as sources to further in-depth information.
This paper is organized as follows. Section~\ref{FRIBscience}
summarizes the science case for FRIB, in the context of the overarching
science questions. The interdisciplinary aspects of the FRIB program
are presented in Sec.~\ref{interdisc}. Section~\ref{expth} discusses
various ways of enhancing the coupling between experiment and
theory. High performance computing will play a key role in FRIB
science program; this is discussed in
Sec.~\ref{computing}. Sections~\ref{organization} and \ref{sec:talent}
talk about organization of the FRIB theory community and educational
aspects, respectively. Finally, Sec.~\ref{conclusions} offers a broad
perspective on FRIB theory.

\section{FRIB Science Overview}\label{FRIBscience}

The science case for FRIB has been formulated over many years and is
well documented.\cite{Decadal2012,TribbleReport,RISAC} In short, the
facility will address -- on many levels -- the overarching science
questions identified by the U.S. National Academy of Science in the
fourth decadal survey of nuclear physics entitled \emph{Exploring the
  Heart of Matter}:\cite{Decadal2012} \textit{1) How did matter come
  into being and how does it evolve?}  \textit{2) How does subatomic
  matter organize itself and what phenomena emerge?}  \textit{3) Are
  the fundamental interactions that are basic to the structure of
  matter fully understood?}, and \textit{4) How can the knowledge and
  technological progress provided by nuclear physics best be used to
  benefit society?}  Together with experiment, future developments in
low-energy nuclear theory and computational science will be critical
in answering these questions.

\begin{figure}[bh!] 
  \centerline{\includegraphics[width=0.6\textwidth]{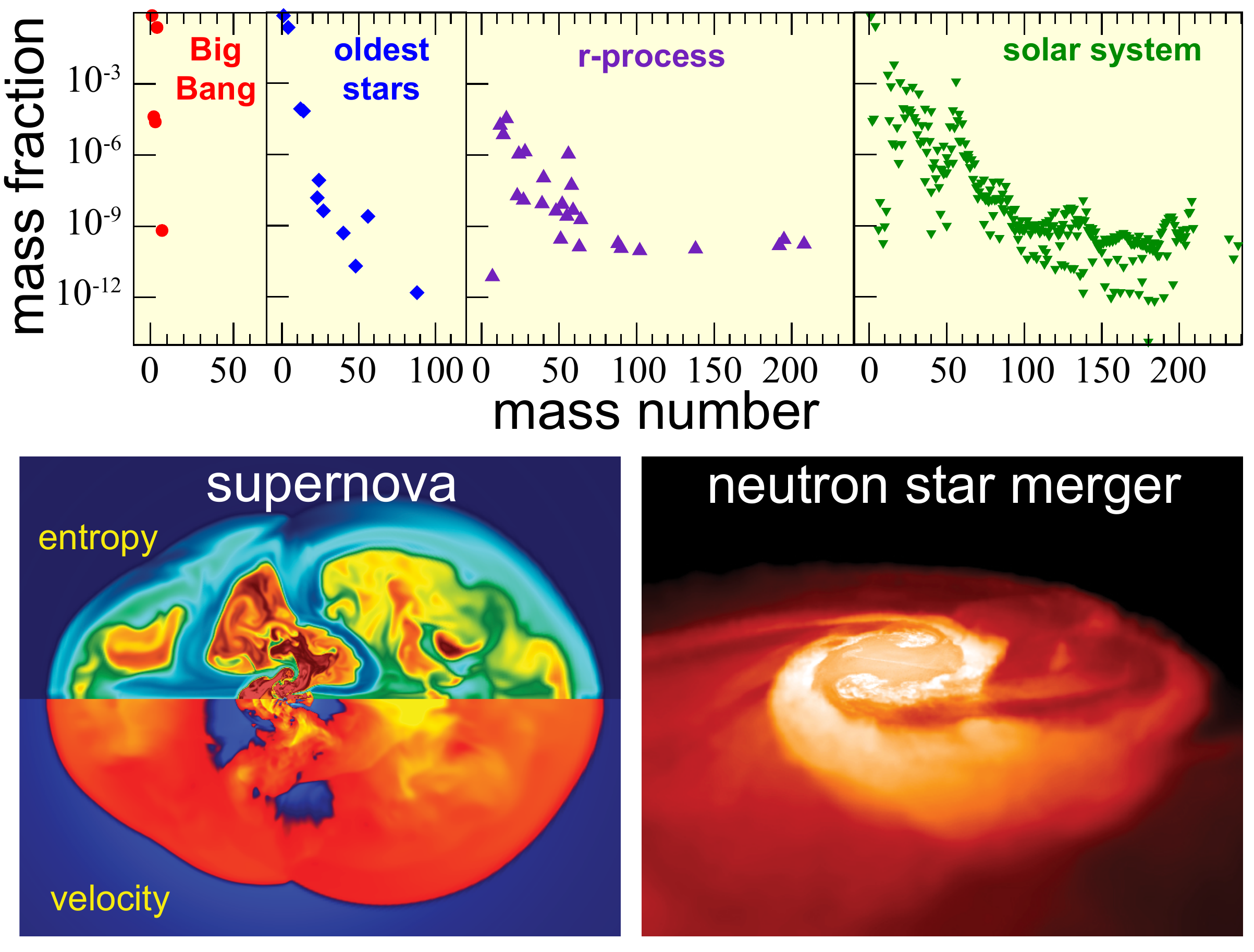}}
  \caption[T]{\label{abundance} Top: Understanding the observed
    sequence of abundance enrichment of
    nuclides\protect\cite{(Sch08a)} is a challenge to theory.  Bottom:
    Advanced simulations of supernova\protect\cite{(Bru13)} (left) and
    neutron star mergers\protect\cite{(Ros13)} (right) - possible
    $r$-process sites.}
\end{figure}

\textbf{The origin of the elements.}  Our radioactive galaxy
demonstrates continuing formation of new short-lived elements by
nuclear reaction sequences (see Fig.~\ref{abundance} and
Refs.~\refcite{(Sch08a),(Wie12)}).  Nuclear 
structure helps dictate galactic chemical evolution. One example is the neutron
driven $s$- and $r$-processes, responsible for building heavy
elements. The resulting final abundances of these processes reflect
nuclear shell structure, which gives rise to the respective
nucleosynthesis paths. Another example concerns the $rp$-process,
which provides a sensitive probe for neutron star surfaces and crusts.
To this day we do not know exactly where the heavy elements were
made. Possible r-process sites include supernovae and neutron star
mergers.  Differences in reaction paths, which depend on the masses
and lifetimes of the nuclei involved, could affect  abundance signatures for
each site.  Theory is a key component to resolve this mystery. Indeed,
nuclear models provide structural input for key nuclei not accessible
to experiment that participate in reaction networks, and large scale
computational simulations -- such as those shown in
Fig.~\ref{abundance}, bottom -- tell us about astrophysical conditions
at possible sites.

The theory roadmap includes deriving nuclear interactions from QCD and
connecting those to the structure of the lightest elements and Big
Bang nucleosynthesis.  The combined effort of new experiments and
theoretical/computational approaches will enable us to accurately
determine all relevant properties and reactions of light nuclei, in
particular neutron-rich nuclei formed during stellar
evolution. Interactions obtained from effective theories of QCD,
density functional theories, experiments, and astrophysical
observations will describe the properties of nucleonic matter found in
nature: the nuclear landscape, neutron stars and supernovae.

\begin{figure}[bh!] 
  \centerline{ \includegraphics[width=0.7\textwidth]{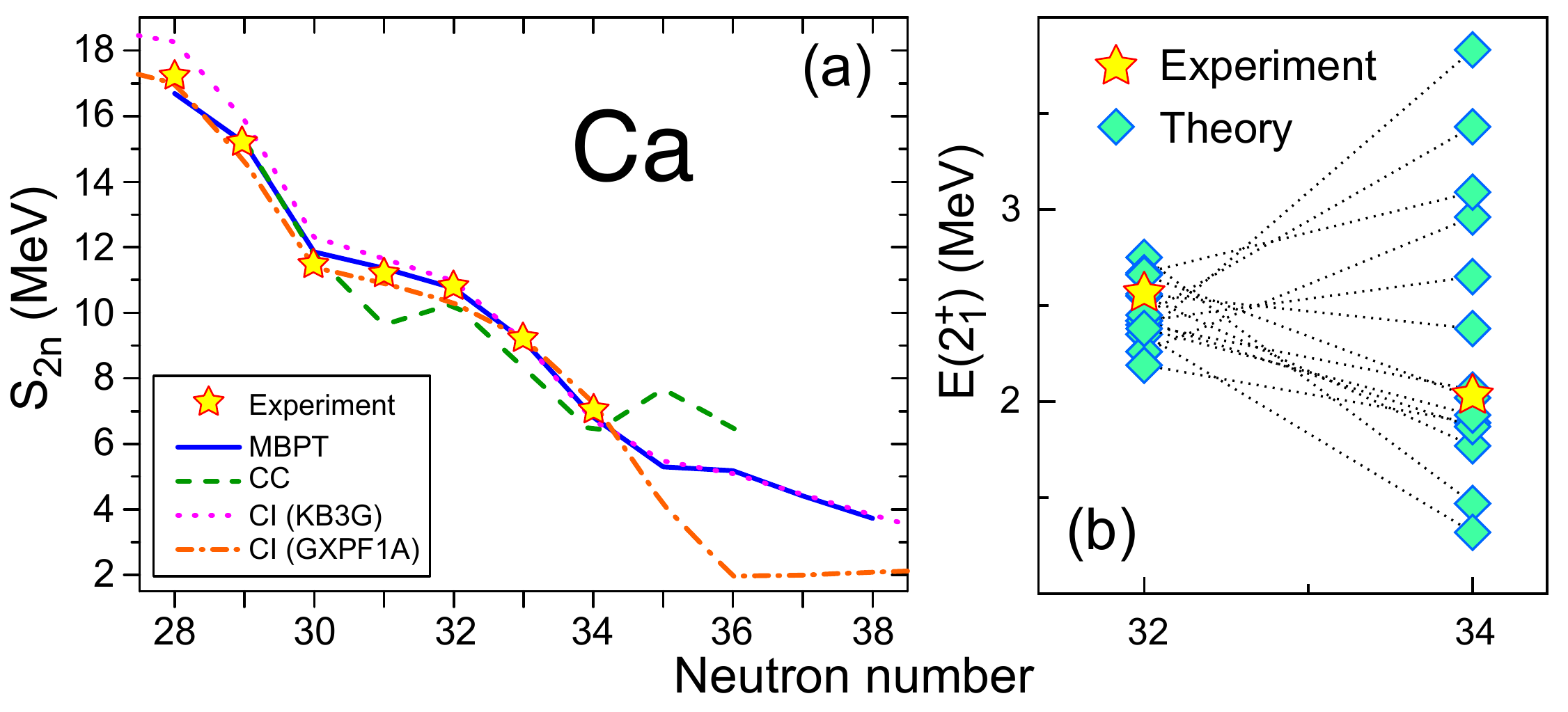} }
  \caption[T]{\label{Ca_chain} Theoretical predictions for (a)
    two-neutron separation energies $S_{2n}$ and (b) $2^+_1$
    excitations (insert) in calcium isotopes compared to experiment
    \protect\cite{(Gal12),(Wie13),(Ste13a)} and in nuclei yet to be
    measured. Theoretical models shown include: Many Body Perturbation
    theory (MBPT), coupled cluster (CC) approach, and configuration
    interaction (CI) methods with different effective interactions
    (cf. Refs.~\protect\refcite{(Wie13),(Ste13a)} for details).  }
\end{figure}

While great progress has been made in the last decade in the
theoretical description of nuclear structure by \textit{ab initio}
methods, configuration interaction approaches, and nuclear density
functional theory, the exploration of neutron-rich systems is still in
its infancy.  Figure~\ref{Ca_chain} provides theory predictions for
the neutron-rich calcium isotopes, which are a frontier for probing
nuclear forces and shell structure.  Predictions for masses (by way of
two-neutron separation energies) show good agreement for measured
nuclei, but diverge where not yet constrained by
experiment.\cite{(Gal12),(Wie13)} This divergence is especially
evident for the $2^+_1$ excitation energies.\cite{(Ste13a)} The
interplay between theory and experiment at FRIB will lead to a robust
phenomenology with controlled and quantitative uncertainties for the
theory predictions of unmeasured nuclei, see Sec.~\ref{expth} for more
discussion.

\textbf{Organization of sub-atomic matter.}  The nature of nuclear
forces and the mechanism of nuclear binding produce amazingly regular
patterns in nuclei. Theory provides the framework to understand the
emergence of these collective phenomena.

Finite nuclei exhibit phase-transitional behavior, critical points as
a function of particle number, spin, and temperature. To understand
what causes the emergent phenomena in atomic nuclei, we need
predictive models of small- and large-amplitude collective motion,
such as those involved in fission and heavy-ion fusion. The atomic
nucleus is an open quantum system\cite{(Mic10)}; hence, phenomena of
nuclear structure are intimately connected to reactions, and both
should be described in a unified way. There are challenges in
achieving this unified description: the inclusion of the particle(s)
continuum and its impact on properties of weakly bound states (such as
halos) and unbound nuclear states, and understanding the role of
reaction thresholds on the appearance of collective cluster states.

\begin{figure}[htb]
 \centerline{\includegraphics[width=0.8\textwidth]{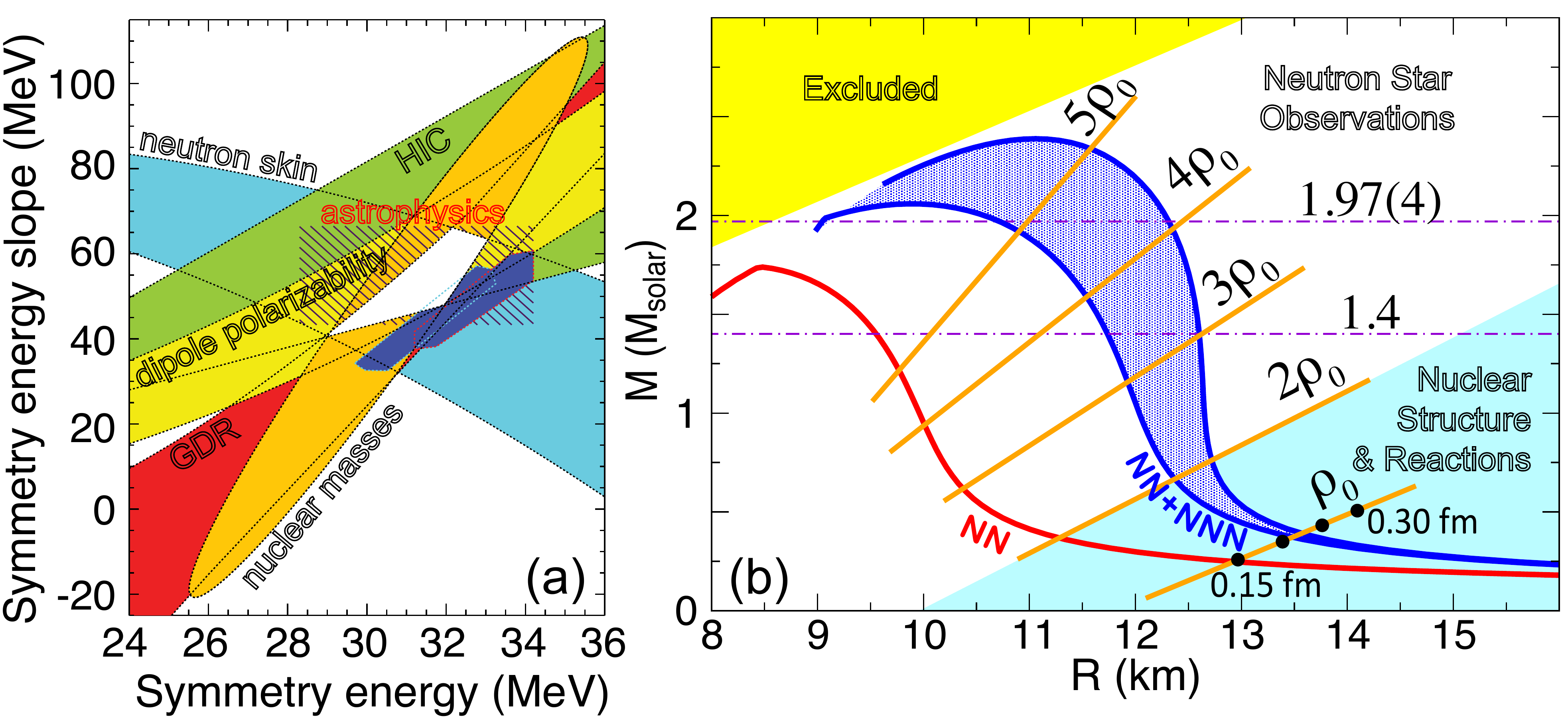}}
\caption{\label{NMatter} Left: Summary of constraints on symmetry
  energy parameters.\protect\cite{(Lat13)} The experimental
  constraints from nuclear data (masses, neutron skin thicknesses,
  dipole polarizability of $^{208}$Pb, giant dipole resonances (GDR),
  and isotope diffusion in heavy-ion collisions (HIC) are marked.  The
  hatched rectangle shows constraints from fitting astrophysical
  observations. The two closed regions show neutron matter
  constraints. The enclosed white area is the experimentally-allowed
  overlap region.  Right: Predicted relation between mass and radius
  of a neutron star modeled with forces involving two nucleons (NN)
  and forces also involving three nucleons
  (NN+NNN).\protect\cite{GandolfiNS} The three-nucleon forces are both
  essential and poorly known, as indicated by a dark blue uncertainty
  band. The orange lines roughly indicate the predicted central
  density of the neutron star. The black dots mark the predicted
  values of the neutron skin in $^{208}$Pb. The accurate measurement
  of a large neutron star mass M=1.97(4) $M_{\rm solar}$ provides a
  strong constraint on theoretical models.  }
\end{figure}

Extended nucleonic matter is another avenue to understand emergent
phenomena. Only with theory can we explore the connection between
neutron-rich matter in the Cosmos and in the laboratory.
Figure~\ref{NMatter} illustrates the multi-disciplinary nature of this
quest.
Answers to many challenging scientific questions ranging from the
dynamics of supernova and heavy-ion collisions to the structure of
neutron stars and rare isotopes, all depend critically on the Equation
of State (EOS) of neutron-rich nucleonic matter. Fig.~\ref{NMatter} demonstrates that the
isospin-asymmetric part of the EOS, namely the density dependence of
nuclear symmetry energy, is still not fully understood.  As seen in
Fig.~\ref{NMatter}(a), while significant effort has been devoted to
constraining the symmetry energy parameters around the saturation
density from the data obtained in terrestrial laboratories,
astrophysical observations, and nuclear theory, large uncertainties
still remain.\cite{(Lat13),Li2013} Of particular importance is the
determination of the symmetry energy at supra-saturation densities.

Figure~\ref{NMatter}(b) (from Ref.~\refcite{GandolfiNS}) displays the
mass-radius relation for a neutron star as predicted by various
theoretical models. The typical mass of a neutron star is about 1.4
solar masses, and the typical radius is thought to be about 12 km. One
of the main science drivers of FRIB is the study of nuclei with
neutron skins three or four times thicker than is currently
possible. Studies of neutron skins in heavy nuclei and investigations
of high-frequency nuclear oscillations and intermediate energy nuclear
reactions with a range of proton and neutron-rich nuclei will help pin
down the behavior of nuclear matter at densities below twice the
typical nuclear density $\rho_0$. At higher densities, relativity and
the observations of a nearly two-solar-mass neutron
stars\cite{(Dem10a),(Ant13)} place severe constraints on the
relationship between the pressure and density of nuclear matter.

\textbf{Fundamental symmetries.}  Experimental tests of the Standard
Model using the nucleus as a laboratory include: searches of atomic
electric dipole moments (EDM) in rare isotopes; parity violation tests
in Fr; CKM matrix unitarity tests by superallowed $\beta$ decay
measurements in $N\approx Z$ nuclei; and searches of exotic scalar and
tensor couplings in $\beta$ decay.
 
Here again, theory and experiment go hand in hand. A variety of
nuclear-structure calculations are critical to tests of the Standard
Model. These include the isospin-mixing corrections in superallowed
$\beta$ decays; nuclear anapole moments in parity violation; Schiff
moments for atomic EDM searches; ordinary and neutrinoless
double-$\beta$ decay matrix elements, and comparison with
observables. These theory predictions are typically needed with high
accuracy and quantified uncertainties. In addition, new weak
interaction signatures need to be explored to probe astrophysical
environments. For example, neutrino signatures probe the burning
conditions and chemistry of the solar core and define the physics of
core collapse supernovae. Electron capture in neutron star crusts
affects neutron star cooling. These studies rely strongly on
theoretical predictions.

\textbf{Nuclear theory and society.}  Last but not least, nuclear
physics can and should be used for the benefit of society. The theory
roadmap includes theoretical advancements relevant for medical
applications and for stockpile stewardship. Examples are an {\em ab
  initio} theory for light-ion fusion, a microscopic theory of
spontaneous and neutron-induced fission, and reaction theory for
medium and heavy nuclei.

Similar to astrophysics, an important stockpile stewardship
application involves a complex network of neutron-induced reactions on
unstable nuclei that cannot be accessed directly in the laboratory due
to the short lifetime of the targets. On the other hand, an indirect
approach leading to the same compound nucleus as the desired reaction,
such as $(d,p)$, which may be performed in inverse kinematics with
radioactive beams at FRIB, can provide important guidance to infer the
reaction of interest.\cite{(Esch12)} However, since this ``surrogate''
reaction may populate slightly different angular momenta and parities
in the entrance channel, a robust theory for nuclear reactions is
needed in order to properly infer the results for the desired
reaction.

Progress made in the understanding of nuclear reactions within an {\em
  ab initio} framework can provide important information to understand
fusion reactions in both astrophysical and terrestrial environments. A
recent example concerns uncertainties in the differential cross
section for elastic $n$-$^3$H scattering, which need to be of the
order 5\% to reliably infer a fuel density for inertially confined
fusion experiments.\cite{(Fre10)} Theoretical
calculations\cite{(Nav11)} based on a reaction theory using the {\em
  ab initio}, no-core shell model and the resonating group
method\cite{(Qua09)} were able to achieve this accuracy and compare
well with data extracted from later experiments.\cite{(Fre11)}
Similar calculations have recently yielded first-principles results
for the $d$($^3$H,$n$)$^4$He fusion reaction\cite{(Nav12)}.

\section{Interdisciplinary Aspects of the FRIB Scientific Program}\label{interdisc}

\begin{figure}[bt!]
  \centerline{\includegraphics[width=0.65\textwidth]{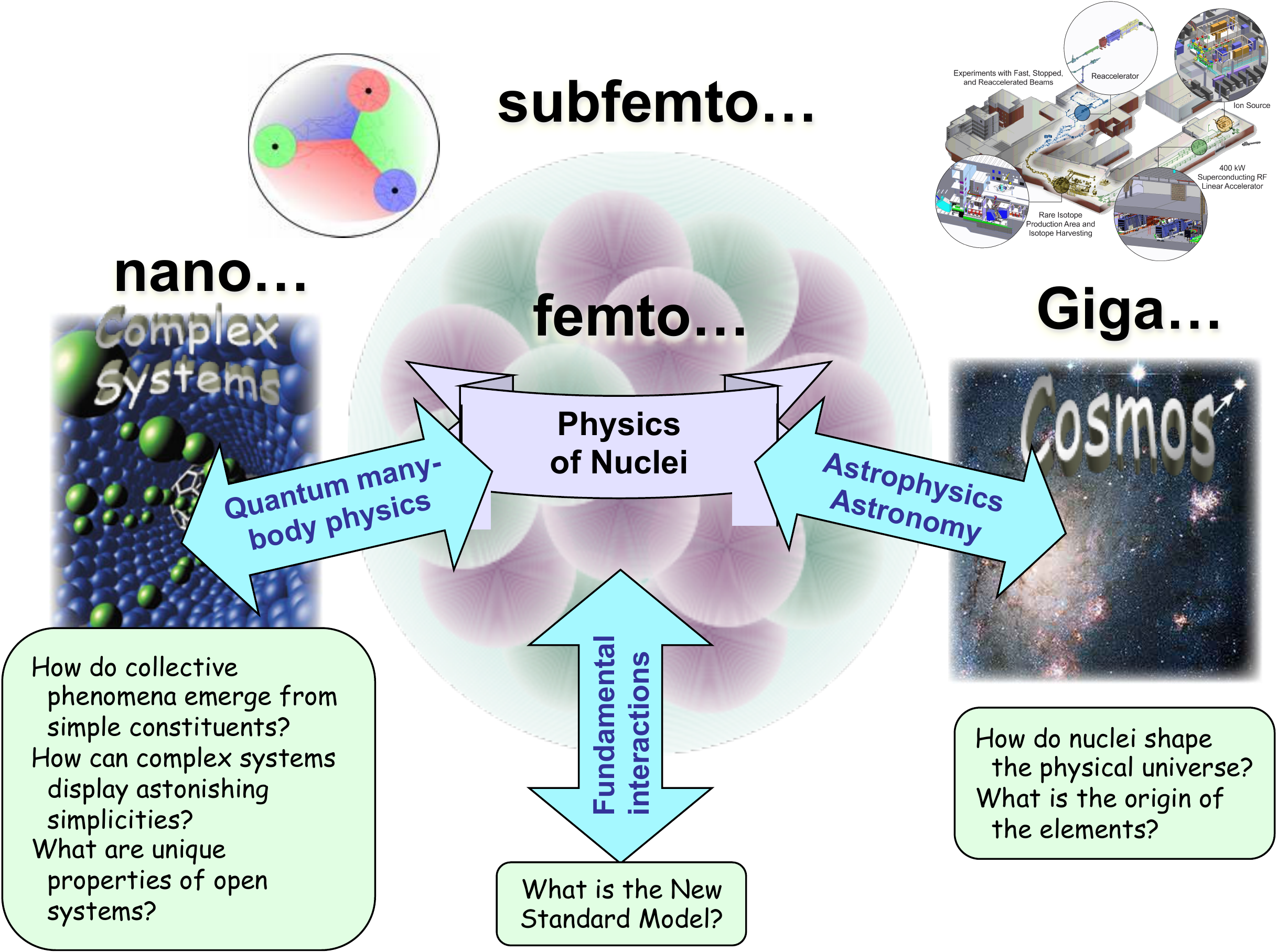}}
  \caption[T]{\label{FRIBintersections} There are diverse
    intersections of FRIB science with research in other sub-fields of
    physics, with shared fundamental questions.}
\end{figure}

Physics with exotic nuclei has intimate connections to many research
areas outside nuclear structure, reactions, and nuclear
astrophysics, see Ref.~\refcite{IntellRIA} and Sec.~5 of
Ref.~\refcite{LRP07}.  Figure~\ref{FRIBintersections} illustrates some
of these intersections and the shared fundamental questions that tie
them together.

A particularly profound synergy exists between research in FRIB
physics and research in astrophysics and cosmology. This synergy also
ties in to neutrino
physics.  The near future will see 30-m class optical telescopes and a
myriad of new observational probes of the cosmos, from measurements of
polarization in the cosmic microwave background (CMB), to 21-cm probes
of high redshift ($z \sim 10\mbox{--}100$), to new X-ray (e.g.,
NuStar) and Gamma-ray (e.g., Fermi) observatories. Nuclear physics,
nuclear astrophysics, and neutrino physics will be key in unraveling
what the new data may mean.

As already emphasized, understanding the origin of the elements around
us is one of the quests of modern science. The answer lies in
understanding the synthesis history of the elements in the universe,
and how this evolution gives us insight into the origin and evolution
of the structures we see, and the nature of dark matter and dark
energy.  New data on the history of nucleosynthesis from quite high
redshift to the present epoch will present nuclear astrophysicists
with exciting opportunities. At issue is whether the history of
nucleosynthesis and star formation is consistent with the picture we
have for the mass assembly history of galaxies. Insight into this
issue may be key to understanding, for example, the origin and
evolution of the dwarf spheroidal galaxies which, in turn, may give
insights into the nature of dark matter.\cite{(Boy12)}

CMB observations have given us, or will give us, precise
determinations of the baryon density of the universe and the ratio of
relativistic to non-relativistic energy density at the epoch of photon
decoupling. These, combined with increasingly precise determinations
of the primordial deuterium and helium abundances, are creating a
nearly over-determined situation for Big Bang Nucleosynthesis
(BBN). This turns BBN into an even more powerful probe of the physics
of the early universe, particularly of new physics in the neutrino
sector, and potentially the QCD epoch. However, there are many open
questions involving the synthesis of $^7$Be and $^7$Li in BBN (with
new physics) and the subsequent fate of these species in stars. These
are central problems in nuclear astrophysics.\cite{(Bal13)}

A quantitative understanding of stellar evolution, stellar explosions,
and the compact objects they produce, relies on nuclear physics. In
the past decade, advances in theory, computation and simulation have
helped interpret diverse astrophysical phenomena, unravelling how
nuclear physics input shapes observable outcomes.  Input from
experimental nuclear and neutrino physics has played an important
role, but theory is essential to access the enormous range of ambient
conditions realized in astrophysics. The nuclear and weak reactions,
the equation of state and transport properties of hot and dense matter
that play a central role in the most extreme conditions are seldom
within reach of direct experiments. A quantitative theory of nuclei
and nuclear matter with quantifiable errors and well understood model
systematics can unravel mechanisms that power supernova, gamma-ray
bursts, x-ray bursts and neutron star mergers and provide fundamental
tests of nuclear physics under extreme conditions. For example, the
density dependence of the nuclear symmetry energy influences the
spectrum and the time structure of the supernova neutrino
signal\cite{(Rob12),(Rob12a),(Mar12a)} and the gravitational wave
signal from binary neutron star mergers\cite{(Bus12)}. The properties
of the neutron star crust, where neutron-rich nuclei and matter at
subnuclear density both play a role, are helping interpret a host of
x-ray phenomena associated with accreting neutron stars, and in some
cases, permitting us to develop a quantitative theory to predict
thermal evolution on timescales of years\cite{(Pag13)} and motivating
a slew of new observations.

There are also numerous profound connections between FRIB science and
many-body physics. Despite the fact that the number of nucleons in
heavy nuclei is small compared to the number of electrons in solids or
atoms in gases, nuclei exhibit emergent phenomena that are present in
other complex systems studied by quantum chemists, atomic, molecular,
and condensed-matter physicists, and materials scientists. Atomic
nuclei exhibit both fundamental and emergent behavior; hence, they
provide important clues to our understanding of the transition from
microscopic to mesoscopic, and to macroscopic.

One example of fruitful interdisciplinary research that bridges
between nano- and femto-scales is the physics of strongly coupled
superfluid systems, such as neutron droplets and cold atoms close to
the unitary regime: both have been successfully treated by many-body
nuclear techniques.\cite{(Gez08),(Zin13),(Ham13)} Another example is
Cooper pairing. Nucleonic superfluidity lies at the heart of nuclear
physics:\cite{(BroZel)} it is present in finite nuclei and in the
nuclear matter of neutron stars, where spatially anisotropic pairing
fields, also discussed in the context of high-temperature
superconductivity in novel materials, are expected.  Theoretical
concepts and tools are shared between the fields. For example
techniques developed in nuclear structure physics, such as random
matrix theory\cite{(Pap07),(Zel10)} and semiclassical
methods\cite{(Mag11)} have been carried over to the study of
mesoscopic systems such as quantum dots and clusters of atoms.

One of the main goals of many-body physics that is shared with nuclear
physics is to understand how collective phenomena emerge from simple
constituents. We know that complex systems can display astonishing
simplicities associated with dynamical many-body symmetries, symmetry
breaking effects, and quantum phase transitions, and the atomic
nucleus shows many examples of collective behavior.  The many-body
behavior of neutrinos in a core-collapse supernova, the only many-body
system driven by weak interactions, reveals an intriguing connection
between neutrinos, emergent properties in many-body physics, and
nucleosynthesis.

Open quantum systems, whose properties are affected by the environment
of decay, scattering, and reaction channels, are also great
interdisciplinary unifiers.\cite{(Mic10)} Many aspects of open quantum
systems that are independent of the system dimensionality are now
explored in atomic nuclei, hadrons, molecules, quantum dots and wires
and other solid-state microdevices, crystals in laser fields, and
microwave cavities. As radioactive nuclear beam experimentation
extends the known nuclear landscape towards the particle drip lines,
the coupling to the continuum space becomes increasingly more
important.  The novel nuclear approaches developed in this context,
such as the continuum shell model,\cite{(Mic09),(Pap13)} are now being
applied to studies of other open quantum systems such as coupled
quantum dots or dipole-bound anions.

\section{Enhancing the Feedback Between Experiment and Theory}\label{expth}

The scientific method uses experimentation to assess theoretical
predictions.  Based on experimental data, the theory is modified and
subsequently can be used to guide future measurements. The process is
then repeated, until the theory is able to explain observations, and
experiment is consistent with theoretical predictions. This positive
feedback in the ``experiment-theory-experiment-" loop can be enhanced
if statistical methods and scientific computing are applied to
determine the couplings between model parameters, parameter
uncertainties, and the errors of calculated observables.

Nuclei communicate with us through observables revealed by
experiment. Some observables are easy to measure; some take considerable effort
and experimental ingenuity. Often, the observable, such as the cross section for a certain nuclear reaction channel, can be used to deduce a quantity of interest, like the distribution of neutron matter in the nucleus or the resonance width. A challenge for FRIB science is that in most cases the extraction of structural data will be model dependent. The reliability of the extracted information  will depend critically on the ability of theory to accurately describe those reaction processes.
Microscopic  approaches to reaction theory consistent with the state-of-the-art structure models are necessary to  reduce the ambiguity of current highly phenomenological models. Statistical tools will be useful  in providing uncertainty quantification. A number of new approaches are being developed for this purpose, but many challenges remain

With nearly 7000 possible isotopes and thousands of pieces of information for each one, not every observable has the potential to impact our understanding in the same way: some measurements are likely to be more important than others. Identification of the key measurements is one way in which theory can provide the foundation for an optimized experimental program at FRIB. By studying the theoretical relevance of the anticipated experimental outcomes, a theoretical assessment of the scientific impact of experiments will help identify critical measurements.  However, we also recognize that as with any new facility, surprises not anticipated by current theory are likely to arise. Theory working hand-in-hand with experiment will be key to unravel the  puzzles. Past examples of this interplay are the unexpected evolution of shell structure with $N$ and $Z$, and the existence of nuclei with large neutron halos and skins.  Often these surprises have provided crucial clues for new physics that was missing from our models.

Theory can also evaluate whether the anticipated experimental errors
are adequate to provide meaningful guidance for further model
developments. Theory should provide input for planning future
experiments by isolating those experimental data crucial to better
constrain nuclear models and validating and verifying model-based
extrapolations.\cite{(Kor13),(Pie12)} An FRIB theory effort, working
closely with the experimentalists associated with FRIB, will serve as
a focal point for facilitating interactions aimed at enhancing the
experiment-theory cycle in the scientific method.

\begin{figure}[tb!]
 \centerline{\includegraphics[width=0.6\textwidth]{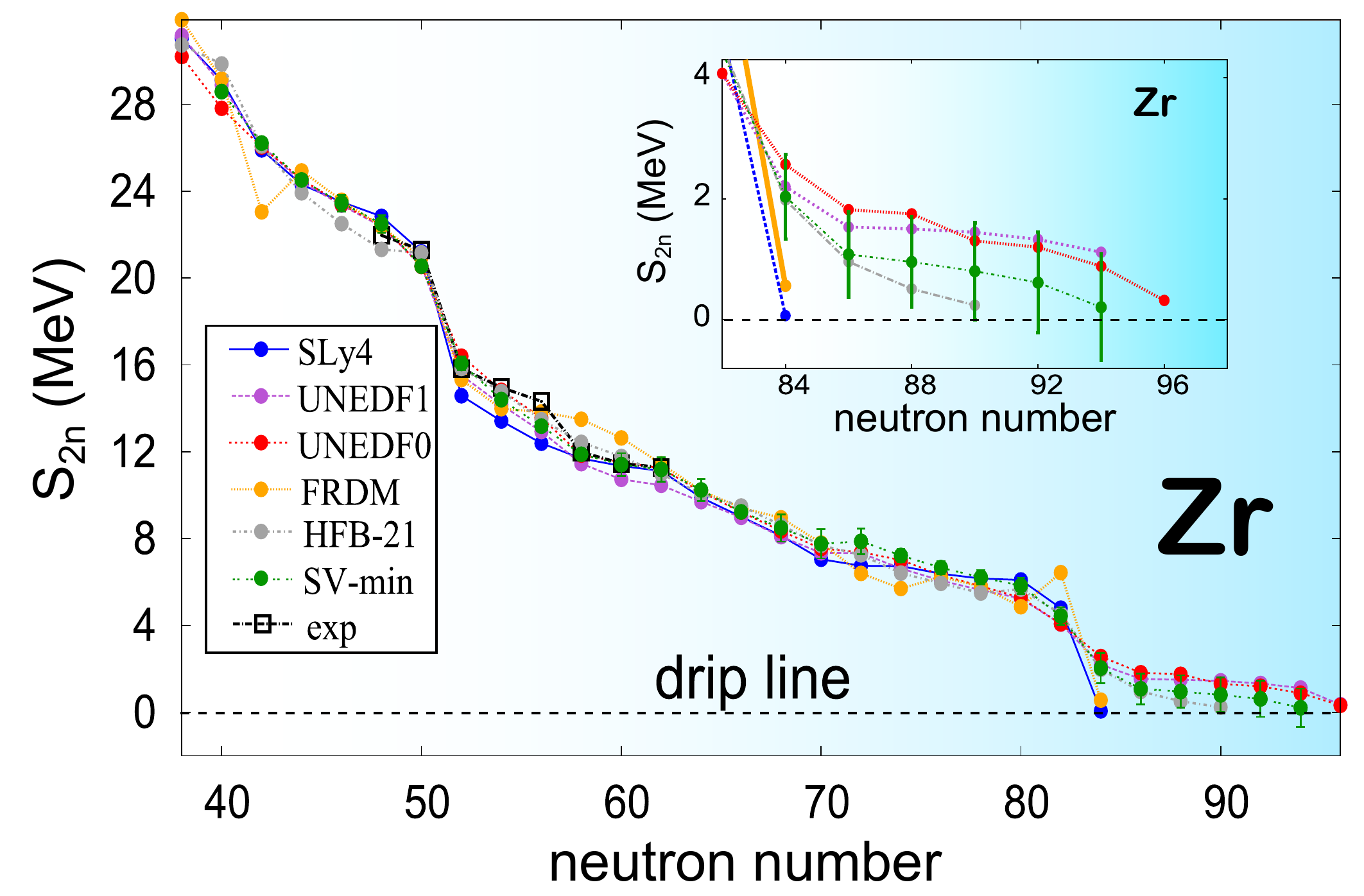}}
\caption{\label{fig:Zr} Calculated and experimental two-neutron
  separation energies of even-even zirconium isotopes.  Calculations
  performed in Ref.~\protect\refcite{(Erl12)} using SLy4, SV-min,
  UNEDF0, and UNEDF1 energy density functionals are compared to
  experiment and FRDM\protect\cite{(Mol95)} and
  HFB-21\protect\cite{(Gor10)} mass models. The differences between
  model predictions are small in the region where data (marked by
  squares) exist and grow steadily when extrapolating towards the
  two-neutron drip line ($S_{2n}=0$). The bars on the SV-min results
  indicate statistical errors due to uncertainty in the coupling
  constants of the functional. Detailed predictions around $S_{2n}=0$
  are illustrated in the inset.  }
\end{figure}

By taking advantage of high-performance computing, nuclear theory is
developing tools to deliver uncertainty quantification and error
analysis for theoretical studies. Statistical tools can also be used
to assess the information content of an observable with respect to
current theoretical models. Such technologies are essential for
providing predictive capability, to estimate uncertainties, and to
assess extrapolations -- as theoretical models are often applied to
entirely new nuclear systems and conditions that are not accessible to
experiment. As already discussed in Fig.~\ref{Ca_chain}, current
nuclear models do not give consistent answers when going outside
``safe" regions explored experimentally.  Figure~\ref{fig:Zr}
illustrates the difficulties encountered with theoretical
extrapolations towards drip lines. Shown are the two-neutron
separation energies $S_{2n}$ for the isotopic chain of even-even
zirconium isotopes predicted with different theoretical models. In the
region for which experimental data are available, all models agree and
reproduce the data equally well. However, the discrepancy between
various predictions steadily grows when moving away from the region of
known nuclei, because the dependence of the effective force on the
neutron excess is poorly determined.  In the example considered, the
neutron drip line is predicted to be between $N=84$ (FRDM and SLy4)
and $N=96$ (UNEDF0), i.e., the model (systematic) error is
appreciable.  In addition to systematic errors, calculated observables
are also subject to statistical errors due to uncertainties in model
parameters.\cite{(Klu09),Rei10,(Fat11)} Figure~\ref{fig:Zr} indicates
that the statistical error predicted with the SV-min energy density
functional gradually grows with $N$. This is primarily caused by the
isovector properties of the model that are not well constrained by the
current data. For other examples, see
Refs.~\refcite{(Kor13),(Pie12),(Afa13),(Erl13)}.  Experimentally, FRIB
with its extended reach has the possibility to produce $^{124}$Zr and
determine its two-neutron separation energy. The figure illustrates
that, with the theoretical understanding of the origin of model
uncertainties coupled with the new data, a significant improvement in
the precision of nuclear models will be possible.

\section{High Performance Computing Aspects of FRIB Science}\label{computing}

As eloquently stated in the recent decadal survey,\cite{Decadal2012}
``High performance computing provides answers to questions that
neither experiment nor analytic theory can address; hence, it becomes
a third leg supporting the field of nuclear physics."  Large-scale
nuclear physics computations dramatically increase our understanding
of nuclear structure and reactions and the properties of nucleonic
matter.  A series of workshops on computational physics and forefront
areas of nuclear science,\cite{ASCR} including QCD, nuclear structure
and reactions, and nuclear astrophysics, established the importance
and breadth of computational nuclear physics.  A large fraction of the
discussion revolved around topics critical to FRIB science. The
importance of computational nuclear physics has been re-emphasized
recently in influential reports.\cite{Decadal2012,TribbleReport}
Reaching the full potential of the FRIB research program requires
comprehensive investigations of many questions that can only be
addressed using world-leading computational facilities.

\begin{figure}[tb!]
 \centerline{\includegraphics[width=0.75\textwidth]{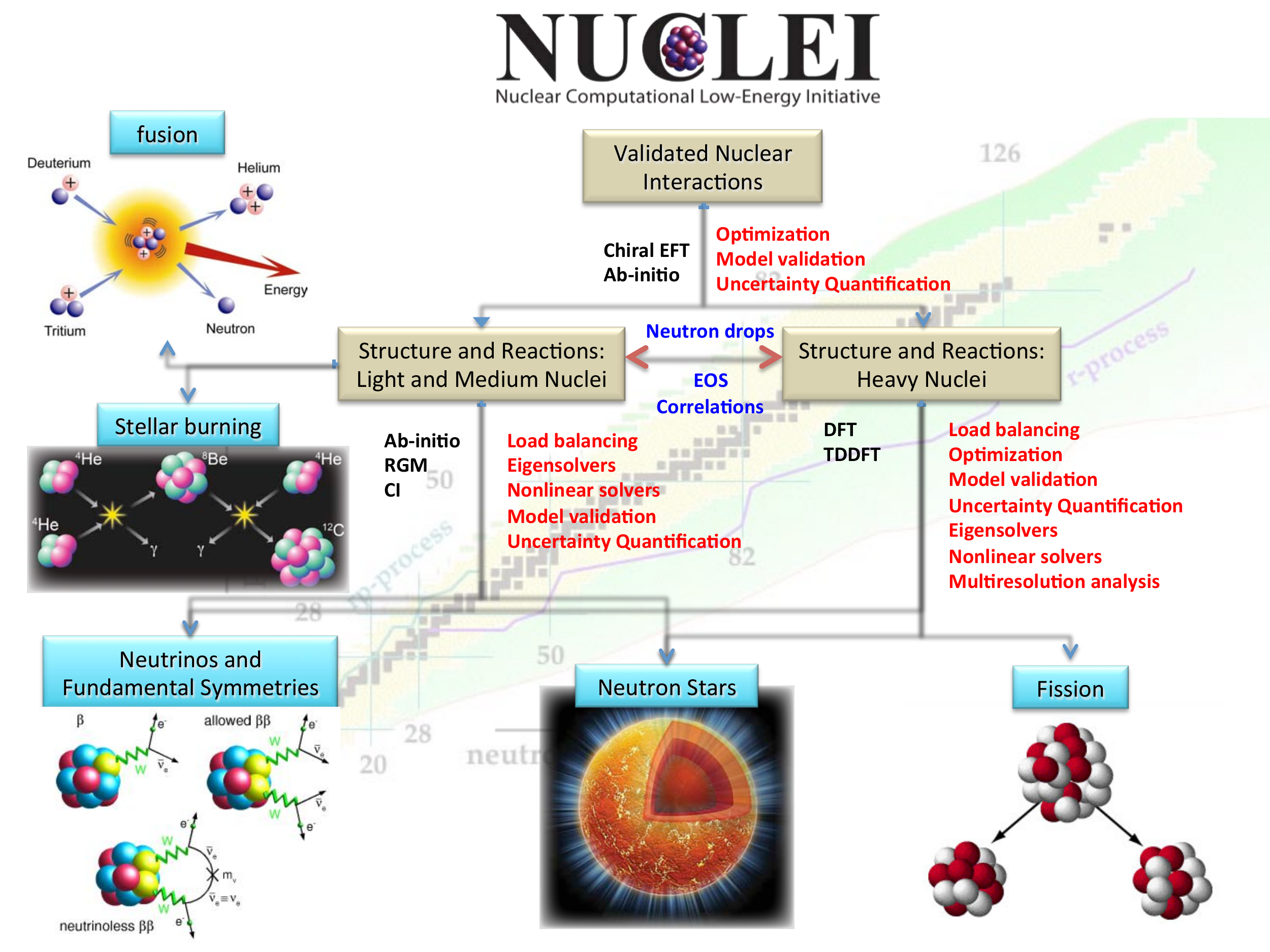}}
\caption{A schematic diagram of the NUCLEI SciDAC
  project\protect\cite{NUCLEI} showing major efforts in nuclear
  physics and their ties to applied mathematics and computer
  science. The NUCLEI project is tied closely to the priorities of the
  experimental nuclear physics program, including particularly FRIB,
  but is also important for neutrinoless double beta decay experiments
  including MAJORANA, EXO, and also to the nuclear physics program
  with energetic electrons at Jefferson
  Lab.\label{fig:nucleistructure}}
\end{figure}

Computational nuclear structure and reactions in the U.S. has advanced
significantly through the UNEDF\cite{UNEDF,FurNPN,bogner2013} SciDAC
project and its successor NUCLEI.\cite{NUCLEI} Both projects joined
forces of nuclear theorists, computer scientists and applied
mathematicians to break analytic, algorithmic, and computational
barriers in low-energy nuclear theory.  Integral to both projects has
been the greatly enhanced degree of quality control: verification of
methods and codes, the estimation of uncertainties, and
assessment. The UNEDF project helped form a coherent nuclear theory
community, opened up new capabilities, fostered transformative science
resulting in high-visibility publications, and advanced the careers of
many junior scientists. The NUCLEI project encompasses significant
components of computational physics relevant to FRIB science; it
bridges the scales from hadronic interactions to the structure and
dynamics of heavy nuclei to neutron stars within a coherent framework.
Figure~\ref{fig:nucleistructure} shows the structure of this project
and its ties to applied mathematics and computer science.

The mathematics and computer science components in NUCLEI are directly
tied to relevant parts of the NUCLEI project.  These groups do
forefront research in applied math and computer science that will be
both broadly applicable across different fields of science and
immediately beneficial to the physics projects in NUCLEI.  Examples of
such joint projects\cite{bogner2013} include: development of the ADLB
(Asynchronous Dynamic Load Balancing) library, which provides scalable
load balancing services for Quantum Monte Carlo calculations on the
largest machines available; new schemes for sparse matrix-vector
operations and development of efficient scalable iterative
eigensolvers for distributed multi-core platforms and topology-aware
mapping of computational tasks to reduce communication overhead
employed in state-of-the-art CI calculations; applications of the
wavelet-based MADNESS (Multiresolution Adaptive Numerical Environment
for Scientific Simulations) to nuclear DFT; derivative-free
multi-parameter optimizations of chiral interactions and energy
density functionals with Practical Optimization Using No DERivatives
(POUNDERS) framework; and uncertainty quantification using the
Gaussian Process Models for Simulation Analysis tool, which is
important in many applications from nuclear interaction input to the
significance of predictions to upcoming experiments.

Many scientific advances in the field were made possible by a rapid
increase in our ability to use the largest-scale computational
resources. Large-scale usage in nuclear structure and reactions in the
U.S.  rose from 80M core-hr/year in 2009 to 320M core-hr/year in
2013. In the future, our ability to use the largest-scale computers
will require additional investments in manpower, particularly as we
transition to new architectures.  These new positions are critical to
enable the effort to scale to the largest-scale machines heading
toward exascale, and to support the FRIB-related experimental programs
in a timely manner.

There are also deep connections between NUCLEI and computational
efforts in neighboring areas. At the shortest-scales, lattice gauge
theory plays an important role, particularly for aspects such as the
three-neutron force that are difficult to isolate, and often access, experimentally.
These can play a very important role in neutron-rich nuclei and
matter.  Simulations of neutron star structure and explosive events on
neutron stars and neutron star mergers will yield critical information
about dense matter, as decribed in Sec. \ref{FRIBscience}.  Supernovae
simulations encompass important studies of radiation hydrodynamics
that illuminate the explosion mechanism and nucleosynthesis, to
studies of neutrino propagation that can tell us more about the
astrophysics of supernovae as well as neutrino properties. Examples of
such simulations are displayed in Fig.~\ref{abundance}.  Every one of
the research areas displayed in Figs. \ref{FRIBrange}-\ref{NMatter}
and \ref{fig:Zr} require state-of-the art computational facilities and
teams of physicists, computer scientists, and applied mathematicians
working together to advance our understanding and to fully exploit the
capabilities of FRIB.

\section{FRIB Theoretical Science Organization}\label{organization}

The current nuclear theory effort in the U.S. related to FRIB is quite
broad and dispersed, with large groups at National Laboratories and
some universities, but also many small university groups.  There are
theorists studying astrophysical phenomena, nuclear reactions, nuclear
equations of state, nuclear structure, and nuclear applications in
national security and isotope R\&D.  Some function largely
independently while others collaborate in large research projects such
as the NUCLEI effort discussed in Sec.~\ref{computing} or topical
collaborations in nuclear theory on neutrinos and nucleosynthesis in
hot and dense matter\cite{NuN} and reactions for unstable
isotopes.\cite{TORUS}

The theory community interested in FRIB physics formed, over a decade
ago, the RIA (``Rare Isotope Accelerator'') and FRIB theory users
groups, whose primary purpose is to identify and prioritize the most
important theory developments in relation to RIA and FRIB projects, to
advocate for the science of radioactive beams, and to be a voice for
the low-energy nuclear theory and astrophysics community.  The members
of the FRIB Theory Users Group regularly join the annual low-energy
nuclear physics community meetings. During the last 10 years, the
theory community has produced two important documents: the RIA Theory
Bluebook (2005)\cite{riatheory1} and an FRIB Theory Users Group Report
(2011).\cite{FRIBTUG11} The former report outlined various scientific
directions necessary for an impactful theory effort, while the latter
focused on issues surrounding education and training of the next
generation of theorists, and included results from a survey on the
needs of the field. While the 2011 survey is new from a time
perspective, we note that the NSAC theory report from
2003\cite{NSACTH} also recommended ways to address pipeline issues.

To address the concerns discussed in the 2011 report, the pipeline of
young theorists must be maintained. In particular, post-doctoral
appointments are crucial for R\&D career development. An national FRIB Theory Fellow program would address the problem, creating
opportunities for young post-doctoral fellows to mature and fostering
continuous interactions among theorists and experimentalists.  The
need for faculty or laboratory positions in nuclear theory has also
been noted in the past, and was the second concern documented in
Ref.~\refcite{FRIBTUG11}.  To overcome this problem, a bridging
program for young faculty could be developed enabling needed growth
into areas of critical need.

The nuclear theory community in the U.S. has also extended beyond its
boundaries.  The report from a recent comparative research review of
nuclear physics for the U.S.\ Department of Energy\cite{DOEcompar}
states: ``New RIB facilities are under construction in Canada, France,
and Asia. With the wider spread of world-leading experimental
facilities, international networking will become more important in the
future. As a first step, DOE has created exchange programs for nuclear
theorists with selected countries.  These programs should be expanded
and include joint graduate education with international partners.''
There are a number of initiatives already in place, that foment
collaborations between countries, namely JUSTIPEN\cite{JUSTIPEN},
FUSTIPEN\cite{FUSTIPEN}, and CUSTIPEN\cite{CUSTIPEN}, which are theory
exchange programs. In addition, International Collaborations in
Nuclear Theory\cite{ICNT} coordinates theory topical programs between
NSCL/FRIB, GSI and RIKEN, to address theoretical issues relevant to
those laboratories.  Finally, the TALENT initiative in graduate
education, discussed in Sec.~\ref{sec:talent} below, has also spurted
from an international framework.

\section{Education  Challenges}\label{sec:talent}

The theory effort around FRIB will also play an important role in the
development of a broad, modern, and attractive educational curriculum
addressing the nuclear many-body problem and related areas.  A
thorough knowledge of up-to-date theoretical methods and phenomenology
will be required to tackle the theoretical and experimental challenges
that will be faced by the next generation of nuclear physicists
working in FRIB science.  However, as discussed in
Sec.~\ref{organization}, most university low-energy nuclear theory
groups are small and, therefore, unable to offer a broad spectrum of
advanced research-based nuclear physics courses.  Fortunately, with
advances in modern educational and computational tools, we are in a
situation where globally coordinated efforts can make a significant
qualitative difference in the way nuclear physics students are
educated.

Recently, nuclear physicists in North America and Europe have teamed
up to launch an educational initiative dubbed Nuclear TALENT (Training
in Advanced Low-Energy Nuclear Theory).\cite{Talent} The long-term
vision of TALENT is to develop a coherent graduate curriculum that
will provide the foundations for a cross-cutting low-energy nuclear
theory research program, and will link modern theoretical approaches with on-going
experimental efforts.  To meet these objectives, educational modules
are being commissioned from the best teachers and specialists in
low-energy nuclear theory. The resulting unique material is being
collected in the form of web-based courses, books, and other
educational resources.  The development of such a knowledge base will
allow highly specialized university groups to benefit greatly from a
broad selection of topics taught by world-leading experts.

In its initial phase, TALENT has selected several topics in low-energy
nuclear physics for teaching modules.  The range of courses is broad,
from nuclear forces and {\em ab initio} approaches, to the theory of
complex nuclei, to nuclear reactions and open quantum systems, to
nuclear astrophysics, to computational tools for nuclear physics.
Some of the topics have already been taught as intensive three-week
courses hosted by the European Centre for Theoretical Studies in
Nuclear Physics and Related Areas in Trento, the National Institute
for Nuclear Theory in Seattle, and GANIL in Caen.  Other courses are
in the pipeline, including one at the Joint Institute for Nuclear
Astrophysics.

One of the major challenges facing TALENT is to develop a robust model
for funding.  To this point, all courses have been run on a voluntary
basis, with financial support for students (lodging and local
expenses) provided by the hosting institutions.  Obviously, to put
this educational initiative on solid ground, sustainable funding is
needed. In addition, a proper model for transferring academic credit
must be developed.  There exist examples of educational collaborations
between various universities in the U.S. where bilateral agreements
have been developed between various colleges of natural science,
enabling an economically sustainable model for credit transfers.  With
such bilateral agreements, TALENT courses can be included in course
curricula of participating institutions, and teaching duties could
also be transferred.  With an FRIB theory center on the horizon, there
are good prospects for a better coordination of educational efforts in
advanced theory of nuclei and nucleonic matter.  While the current
TALENT effort is built around theoretical and experimental activities
in North America and Europe, its expansion into other regions is
envisioned.  In particular, considering the scale of experimental
efforts in rare-isotope science in Japan and China, it is
anticipated and hoped that Asia will soon join the initiative.

\section{Summary}\label{conclusions}
An understanding of the properties of atomic nuclei and their
reactions is essential for a complete description of nuclei, an
explanation of element formation and the properties of stars, and for
present and future energy, defense, and security applications. This
requires a coherent picture across many energy scales, all the way
from the interactions between nucleons to the superheavy nuclei and
neutron stars.

The roadmap for low-energy nuclear theory is well
established.\cite{UNEDF,FurNPN,riatheory1} It involves the extension
of {\em ab initio} and configuration interaction approaches all the
way to medium-heavy nuclei, and the quest for a universal nuclear
density functional that will allow description of all nuclei up to the
heaviest elements and neutron stars. 
It also includes developments in reaction theory required for a meaningful link to experiment. 
The direct coupling from nucleon-nucleon interaction scales ($\sim$100 MeV) to nuclear binding
scales ($\sim$1--10 MeV) to collective excitation scales ($<$1 MeV),
facilitated by effective field theory, provides a coherent picture of
the structure and dynamics of all nuclei and nucleonic matter found in
astrophysical environments. To realize this vision, the properties of
rare isotopes are an essential guide.

The development of a theoretical framework that connects the light and
heavy nuclei, proton-rich and neutron-rich rare isotopes, and dense
neutron matter is within reach. In the next decade, these developments
will have profound implications for nuclear physics, nuclear
astrophysics, and neighboring areas, such as high-energy astrophysics,
fundamental interaction physics, and hadron structure.

With FRIB, the field has a clear path to achieve its overall
scientific goals and to answer the overarching questions. With FRIB,
we will have the ability to produce the key isotopes now
unavailable. FRIB will be the world's most powerful facility to
explore the rare-isotope frontier, making nearly 80\% of the isotopes
predicted to exist for elements up to uranium and providing access to
beams of the most interesting isotopes. By taking advantage of the
unique coupling of crucial data from FRIB and other radioactive beam
facilities with advanced theoretical frameworks and high-performance
computing, nuclear theory will be able to develop a predictive picture
of nucleonic matter. This is an exciting perspective.

\section*{Acknowledgments}

This work was supported by the U.S. Department of Energy under
Contract No.  DE-FG02-96ER40963 (University of Tennessee),
Nos.~DE-SC0008499/DE-SC0008533 (NUCLEI SciDAC Collaboration),
DE-AC02-06CH11357 (Argonne National Laboratory), DE-AC52-07NA27344
(Lawrence Livermore National Laboratory), DE-AC52-06NA25396 (Los
Alamos National Laboratory), DE-SC0004087 and DE-FG52-08NA28552 (Michigan State Univ), and 
DE-AC05-00OR22725 (Oak Ridge National Laboratory); the National Science Foundation under Grant
No.~PHY-1306250 (Ohio State University), PHY-1068022 (Texas A\&M
University-Commerce), PHY-1068271 (Michigan State Univ),
PHY-0970064 and PHY-1307372 (University of California, San Diego),
PHY-1205024 (University of Wisconsin-Madison);
and the National Aeronautics and Space Administration under grant
NNX11AC41G issued through the Science Mission Directorate (Texas A\&M
University-Commerce).


\end{document}